\documentclass{article}
\usepackage[english]{babel}
\usepackage[margin=1in]{geometry}
\usepackage{background}
\usepackage{graphicx}
\usepackage{float}
\usepackage{cite}

\usepackage{hyperref}
\hypersetup{
    colorlinks=true,
    linkcolor=blue,
    filecolor=magenta,      
    urlcolor=cyan,
    pdftitle={Overleaf Example},
    pdfpagemode=FullScreen,
    }

\usepackage{amsmath,array,graphicx}
\usepackage{verbatim}
\usepackage[font=small,labelfont=bf]{caption}

\usepackage[math]{blindtext}

\SetBgScale{4}
\SetBgColor{gray}
\SetBgAngle{90}
\SetBgContents{arXiv: some other text goes here}
\SetBgPosition{-1.5,-10}

\begin{document}

{\centering

{\bfseries\Large Machine learning in front of statistical methods for prediction spread SARS-CoV-2 in Colombia \bigskip}

   {\vspace{0.4cm} \itshape
\normalfont{
    A. Estupiñán\textsuperscript{\dag 1};
    J. Acuña\textsuperscript{1};
    A. Rodriguez\textsuperscript{2};
    A. Ayala\textsuperscript{2};
    C. Estupiñán\textsuperscript{3};
    Ramon E. R. Gonzalez\textsuperscript{4};
    D. A. Triana-Camacho\textsuperscript{5};
    K. L. Cristiano-Rodríguez\textsuperscript{5} and Carlos Andrés Collazos Morales\textsuperscript{6} 
}
}
   {\vspace{0.4cm} \itshape

\textsuperscript{1}%
    Universidad de Investigación y Desarrollo (UDI), Bucaramanga, Colombia.\\
\textsuperscript{2}%
    Universidad Autónoma de Bucaramanga (UNAB), Bucaramanga, Colombia.\\
\textsuperscript{3}%
    Universidade Federal de Pernambuco (UFPE), Recife, Brasil.\\
    \textsuperscript{4}%
    Universidade Federal Rural de Pernambuco (UFRPE), Recife, Brasil.\\
\textsuperscript{5}%
    Universidad Industrial de Santander (UIS), Bucaramanga, Colombia.\\
    \textsuperscript{6}%
    Universidad Manuela Beltrán (UMB), Bogotá, Colombia. 

}
   {\vspace{0.4cm} \itshape

\normalfont{$^{\dag}$aestupinan4@udi.edu.co}
\\
\normalfont{September 27, 2022}

}

}

\begin{abstract}
 \noindent An analytical study of the disease COVID-19 in Colombia was carried out using mathematical models such as Susceptible-Exposed-Infectious-Removed (SEIR), Logistic Regression (LR), and a machine learning method called Polynomial Regression Method. Previous analysis has been performed on the daily number of cases, deaths, infected people, and people who were exposed to the virus, all of them in a timeline of 550 days. Moreover, it has made the fitting of infection spread detailing the most efficient and optimal methods with lower propagation error and the presence of statistical biases. Finally, four different prevention scenarios were proposed to evaluate the ratio of each one of the parameters related to the disease.
\bigskip


\end{abstract}

\section{Introduction}
In the last two years, the world has suffered a great situation, in the face of the epidemic known as COVID-19, to date more than three million deaths have been registered, which has caused great concern in the face of the high mortality caused by this lethal disease, especially in older adults. Even today, this disease claims fatalities, attacking people with different age ranges, and different types of mutations of the SARS-CoV-2 virus that cause the COVID-19 disease have been presented \cite{deeks2020antibody,dinnes2021rapid}. Said variants, strains, or mutations of the virus, have resulted in a higher infection rate, the virus being more aggressive both in infection and in the symptoms presented by people infected with said virus.

\noindent In light of the numerous and dangerous consequences that this pandemic has brought to the world \cite{hays2005epidemics,samal2014historical}, many scientists, including biologists, virologists, physicists, and statistical mathematicians, have been working on being able to reproduce and implement an analytical or numerical mathematical model, which can represent the speed of infection or infection of the SARS-CoV-2 virus, from different methods of differential and statistical calculation, which have been previously used to model the evolution of the infection of different viruses, such as the virus of influenza A of the H1N1 subtype (1918 - 1920) that causes the Spanish grippe, the H3N2 influenza virus (1968) that causes the Hong Kong grippe, the HIV virus (Since the 70s) that causes the Acquired Immunodeficiency Syndrome (AIDS), Ebola (Since 1976) causing the Ebola hemorrhagic fever and SARS (2002 - 2003), the latter being 80 \% similar to the one that caused the current COVID-19 pandemic \cite{monto2013influenza,pujadas2020pandemics,leavitt2021pandemics}.

\noindent  In order to be able to predict, forecast, and prepare a contingency plan against a pandemic, the mathematical models capable of representing the evolution of the infection provoked by this disease in a large number of people \cite{kushwaha2020significant,pinter2020covid,lacitignola2021using,kai2020universal} under certain conditions (simulation scenarios). Within these considerations, relevant parameters take account in the simulations, such as The days of social isolation of the individuals to be studied, social distancing, the adoption and execution of biosafety, and self-care measures, such as the use of face masks and continuous hand washing, among others.

\noindent In this manuscript, an analytical and numerical research study was carried out based on the actual infection data of the COVID-19 disease in Colombia. Therefore, where relevant statistical data were taken into account, such as new cases of infection, the total number of infected, number of deaths, and number of recovered, among other types of data categories taken into account in this work.

\noindent This manuscript is divided as follows, in section \ref{sec_2}, the work’s theoretical framework is shown, followed by a brief conceptual and algebraic description regarding the mathematical methods and models; then, in section \ref{sec_3}, the results are shown and carried out an analysis highlighting and emphasizing the findings, observations, and recommendations to take into account, in order to reduce the speed of spread of the SARS-CoV-2 virus. Finally, in section \ref{sec_4} the conclusions, improvements, and recommendations in front of the pandemic studied in this research work are presented.

\section{Theoretical framework}
\label{sec_2}
\subsection{Classic SEIR model}

One of the most used mathematical models to represent the behavior of a pandemic is the SEIR model. Taking into account that each of the acronyms of this mathematical method denotes a variable to simulate, which are: $S = Susceptible$, $E = Exposed$, $I = Infected$ and $R = Removed$. It should be noted that this model is based mainly on the interaction and evaluation of the variables involved in this model through coupled differential equations of the first degree and order.

\begin{figure}[H]
    \centering
    \includegraphics[width=0.90\textwidth]{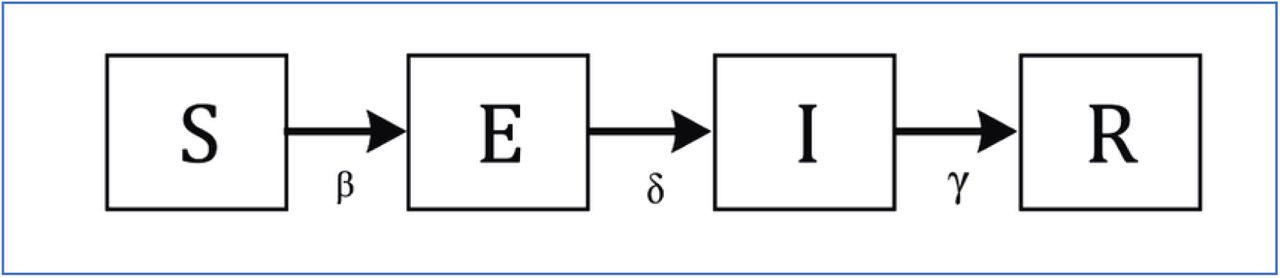}
    \caption{Representation of the relationship between the variables of the SEIR model. Figure extracted of \cite{zisad2021integrated}.}
    \label{fig_1}
\end{figure}

\noindent In Figure \ref{fig_1}, the integration and relationship of the variables $S, E, I$ and $R$ of the basic model is shown, where the $\beta$ is the infectious rate, controls the rate of spread which represents the probability of transmitting disease between a susceptible and an infectious individual, $\delta$, is the rate of latent individuals becoming infectious (average duration of incubation is $1/\delta$) and $\gamma=1/D$ is determined, by the average duration $D$ of infection \cite{zisad2021integrated}.

\noindent The mathematical expressions, which relate the variables and coefficients involved in the classic SEIR model, are the following:
\begin{align}
    \frac{ds}{dt} &= -\beta S(t) I(t) \\
    \frac{dE}{dt} &= \beta S(t) I(t) - \epsilon E(t) \\
    \frac{dI}{dt} &= \epsilon E(t) - \gamma I(t) \\
    \frac{dR}{dt} &= \gamma I(t)
\end{align}

\noindent The set of Coupled Differential Equations (CDE), shown above, denotes the variation in time of each of the simulated variables; for example, for $dS/dt$, reference is made to the number of susceptible patients, possible to change in the time, from an S (Susceptible) state to an E (Exposed) state (See Figure \ref{fig_1}). Similarly, $dE/dt$, $dI/dt$, and $dR/dt$ make reference to the change of states of individuals, in the exposed, infected, and removed compartments, respectively \cite{mahmud2020applying}.

\subsection{Modified SEIR model}

\begin{figure}[H]
    \centering
    \includegraphics[width=0.90\textwidth]{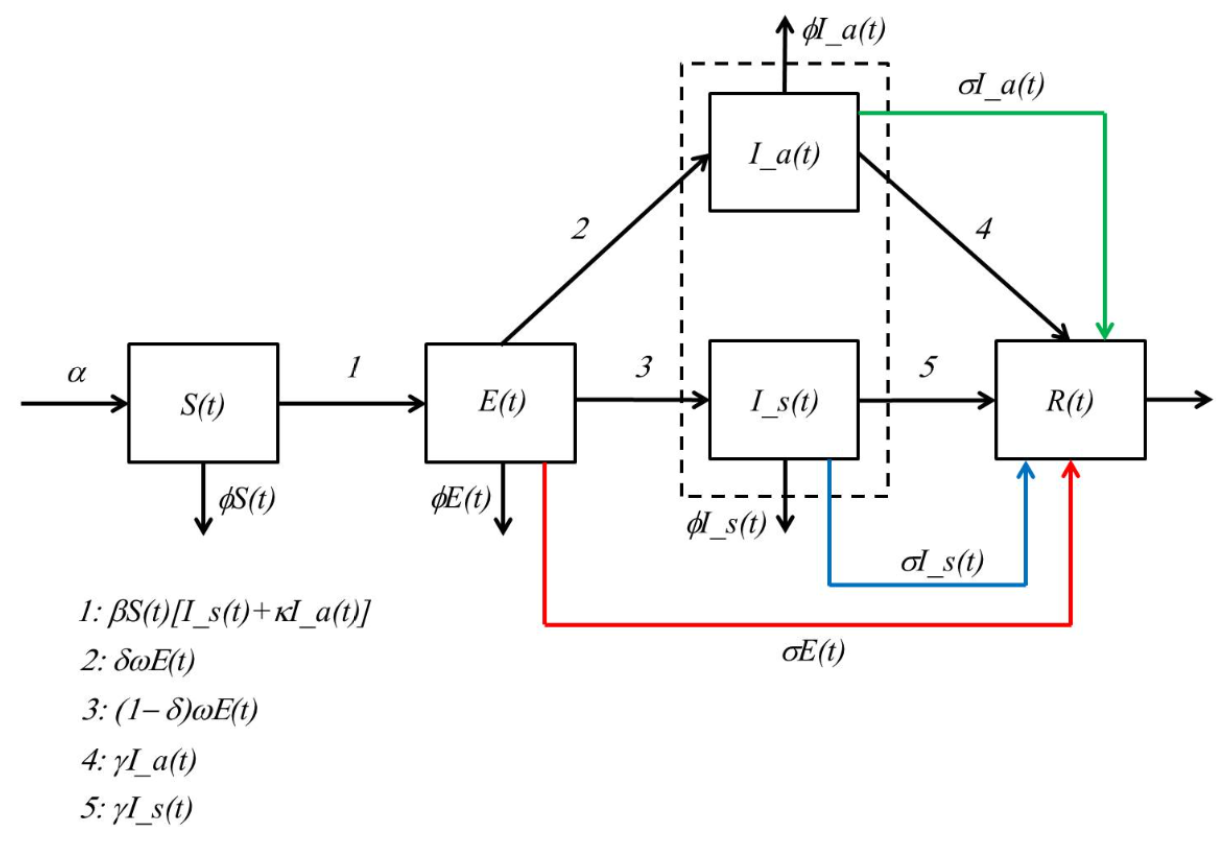}
    \caption{Representative diagram of the interaction of the variables involved in this model. Figure taken from \cite{gonzalez2020adapted}.}
    \label{fig_2}
\end{figure}

In order to obtain a more realistic model of the COVID-19 disease, an adapted or modified model has been taken as a reference, which can better represent the time evolution of the infection. In Figure \ref{fig_2}, you can see the relationship between the simulated variables involved in this model to be implemented \cite{gonzalez2020adapted}.

\noindent This mathematical model represents and simulates different scenarios in which the biosafety measures adopted by the governments in the different countries of the world are taken into account. For example, a scenario in which isolation and quarantine measures are not included can be defined as Scenario 0, and it can be reproduced from the following system of differential equations:

\begin{align}
    \frac{ds}{dt} &= -\alpha -\phi S(t) - \beta S(t) [I_s(t) + \kappa I_a(t)] \\
    \frac{dE}{dt} &= \beta S(t) [I_s(t) + \kappa I_a(t)] - (1 - \delta)\omega E(t) - \delta \omega E(t) - \phi E(t) \\
    \frac{dI_s}{dt} &= (1 - \delta)\omega E(t) - \gamma I_s(t) - \phi I_s(t) \\
    \frac{dI_a}{dt} &= \delta \omega E(t) - \gamma I_a(t) - \phi I_a(t) \\
    \frac{dR}{dt} &= \gamma I_s(t) - \gamma I_a(t) - \phi R(t)
\end{align}

\noindent Unlike the classic SEIR model, herein presented model considers people who appear as infected asymptomatic ($I_a$) and infected symptomatic ($I_s$). In addition, the Removed variable R continues to refer to the number of deceased persons and the number of persons recovered from this disease. 

\noindent In addition, it is essential to take into account that the coefficients that appear in the set of equations shown for "Scenario 0" will have a constant value such as $\alpha$ $=$ Replacement of susceptible individuals rate, $\kappa$ $=$ Multiple of the transmissibility of $I_a$ to $I_s$, $\delta$ $=$ Proportion of asymptomatic infection rate, $\omega$ $=$ Incubation frequency and $\gamma$ $=$ Latency frequency. On the other hand, even for the same scenario, there are coefficients known as control parameters, among which are: $\phi$ $=$ Removal/replacement rate, $\beta$ $=$ Transmission rate and $\sigma$ $=$ Proportion of individuals in quarantine/isolation \cite{gonzalez2020adapted}.

\noindent The basic reproduction rate is taken into account in the presented model too. Such parameter is denoted by the variable $R_0$, which can change depending on the scenario to be simulated, in this way for "Scenario i-th", the expression of the $R_0$ parameter is:

\begin{equation}
    R_0 = \frac{\beta(1+\kappa) + \omega}{2\gamma + i\sigma}
\end{equation}

\noindent In this light, it can be seen that in terms of the $R_0$ parameter, the different scenarios to be reproduced will depend on the characteristic parameters taken ($\beta$, $\kappa$, $\sigma$, and $\omega$). To consider this model's greater detail and explanation, check the reference \cite{gonzalez2020adapted}.

\subsection{Logistic growth model}

Among the many applications that linear differential equations have, this model is applied to represent the population growth of specific individuals, where this population growth will depend mainly on the available natural resources since the number of inhabitants will become smaller as the population size approaches a maximum, imposed by the limited resources of the environment, available to the population studied.

\noindent The equation that models this characteristic type of growth is the following:

\begin{equation}
    \frac{dN}{dt} = r_{max}\frac{(K-N)}{K}N,
\end{equation}

\noindent where $N$ is the size of the population to be analyzed, $r_{max}$ is the base of the population growth rate, and $K$ is the carrying capacity, which represents the maximum size of the population that can support a particular environment.

\subsection{Machine learning polynomial regression method}

The model here proposed for the numerical adjustment allows us to carry out a regression for a polynomial of a particular degree n (for example, linear, quadratic, or cubic), depending on the dispersion of the experimental data to be adjusted with a selected model.

\noindent When it is necessary to model a data set, which has a very considerable curvature, it is necessary to implement the polynomial regression method, which allows being quite sensitive to said changes or curvature modifications in the adjustment of the data set that the user wants to model and adjust.

\noindent By using this regression method, the user can obtain an accuracy up to 7.75 times higher than in the case of fitting with linear regression. The mathematical model used to fit a data set through the polynomial regression method is as follows:

\begin{equation}
    p = \beta_0 + \sum_{i=1}^m \beta_i \cdot  T_i,
\end{equation}

\noindent where $T_i = \Pi_{j=1}^{n} X_{j}^{a_{i,j}}$, $a_{i,j}$ are variable degrees, $a_{i,j} \geq 0$, and $\beta_i$, i = 0,...,n are constants, $\beta_i$, $i>0$. All $T_i$ are refereed to as terms or monomials in $P$. The length of $P$ is $Len(p) = \sum_{i=1}^m \sum_{j=1}^n a_{i,j}$, the size of $P$ is $size(P)=m$ and the degree of $P$ is $Deg(P) = max(\sum_{i=1}^m \sum_{j=1}^n a_{i,j})$. An example polynomial equation is $P = 1.2X^2_1 X_2 + 3.5X_1X^3_2 + 5X_1X_3 +2$. This equation has size 3, degree 4 and length 9 \cite{peckov2012machine}.    

\begin{figure}[H]
    \centering
    \includegraphics[width=0.80\textwidth]{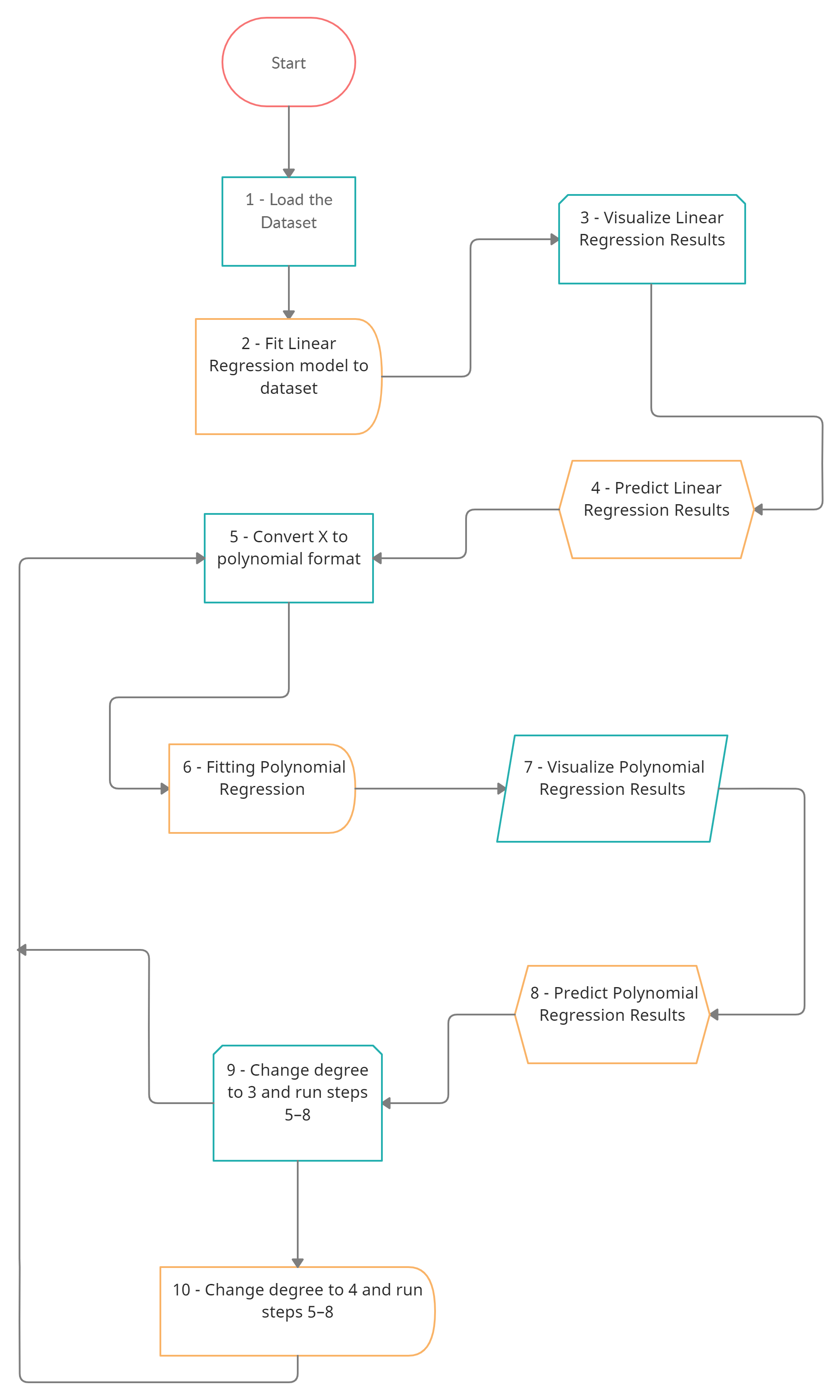}
    \caption{Flowchart, representing the algorithm with the steps; of the numerical method of Polynomial Regression of Machine Learning.}
    \label{fig_3}
\end{figure}

\noindent In Figure \ref{fig_3}, the flowchart of the algorithm of the machine learning method is shown using polynomial regression, whereas the main steps of this method are shown in Figure \ref{fig_3}.

\noindent The first step in executing this method consists of loading the data clean, prepared, and ready to be read by the algorithm. In the second step, a simple linear regression was carried out, which evaluates, compares, and predicts the model with the polynomial regression implemented method. In the third step, the graph is made with the linear regression fit performed on the data to compare said first-order fit. The third step shows the graph with the linear regression fit performed on the data to compare said first-order fit. In the fourth step, some predictions are made for specific points observed in the plotted data distribution to evaluate the method's efficiency for a first linear approximation $(y = b_0 + b_1x)$.

\noindent Step five starts the treatment, transformation, and conversion of matrix x. In which the data to be worked is stored, where now this new matrix will contain the data raised to the maximum degree that we want to adjust, for example of the second order $(y = b_0 + b_1x + b_2x^2)$, this matrix will contain in the first column values of one "1", the second column the values of the data stored in the variable $x$ and the third column the data of the variable $X$ squared. On the other hand, if the polynomial adjustment is of the third order $(y = b_0 + b_1x + b_2x^2 + b_3x^3)$, this matrix will contain in the first column values of one "1", the second column the values of the variable $x$ squared and in the third column, the data of the variable $x$ cubed.

\noindent In step number six, the polynomial regression adjustment process is carried out, a process in which the matrix generated in the previous step is used, in addition to creating the second-degree polynomial regression function. Step seven presents the visualization of the adjustment made using the polynomial regression method. In step number 8, the first efficiency tests of the method are carried out to evaluate the precision accuracy in terms of the prediction of the implemented method for a polynomial of degree 2. Once step number 8 is done, and it is found that the approximation for a second-degree polynomial is not the most desired, we proceed to step number 9, in which we seek to implement a fit of the data with a polynomial function of a higher degree than the one achieved in the previous step, in this case of order three and proceed to repeat the steps from number 5 to 8. Finally, in step number 10, if the polynomial of degree 3 still presents inaccuracies, proceed with the data adjustment using a fourth-order polynomial and repeat the procedure seen in steps 5 to 8.

\section{Results}
\label{sec_3}

\begin{figure}[H]
    \centering
    \includegraphics[width=0.90\textwidth]{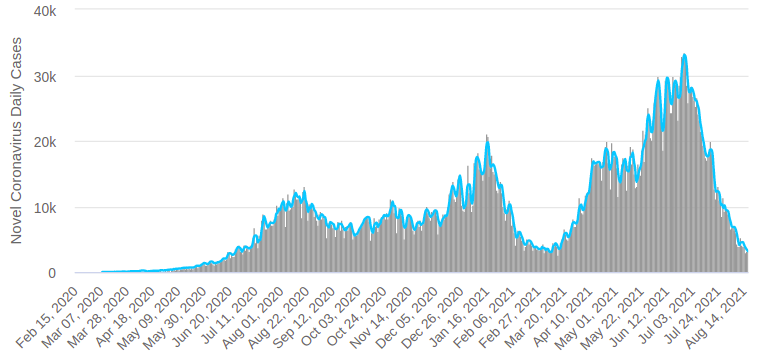}
    \caption{Graph of the new cases of infection by Coronavirus (COVID-19) in Colombia since the beginning of the pandemic in the country. Where the bars of the histogram indicate the number of daily cases and the light blue curve indicates the moving average of cases during 3 days.}
    \label{fig_4}
\end{figure}

Categories can organize the results obtained in this work to analyze the behavior of the studied models and the methods used for each of the different variables analyzed in this research. For this reason, in this section of the analysis of results, we will begin with studying the variable of new cases of infection in Colombia, where a significant variability can be evidenced in both the width and the maximum or peak value reached for each increase in infection during the pandemic (See Figure \ref{fig_4}).

\noindent In Figure \ref{fig_4}, the three highest peaks of infection of the COVID-19 disease can be mainly evidenced on the dates of August 23, 2020, January 19, 2021, and the highest and most recent peak since the beginning of the pandemic, which was present on June 27, 2021. On the other hand, it can be noted how in March, there was a significant decrease in the rate of infection of the disease.

\begin{figure}[H]
    \centering
    \includegraphics[width=0.90\textwidth]{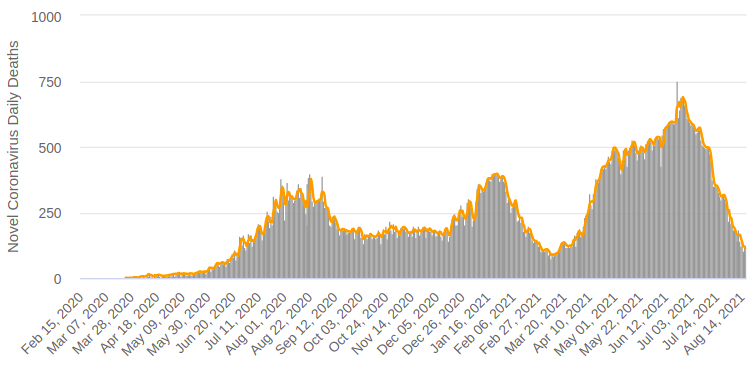}
    \caption{Graph of the data of daily deaths caused by the Coronavirus disease (COVID-19) in Colombia since the beginning of the pandemic in the country. Where the bars of the histogram indicate the number of daily deaths and the dark yellow curve indicates the moving average of deaths during 3 days.}
    \label{fig_5}
\end{figure}

\noindent In addition to the daily infections by Coronavirus (COVID-19), it can be observed that the daily mortality rate due to this disease is a variable that is directly involved with the daily infections and as can be seen in Figure \ref{fig_5}, The behavior of this graph is very similar to the profile seen in Figure \ref{fig_5}. The peaks for the number of deaths by COVID-19 are presented in the same way and for the same dates in Figure \ref{fig_4}, which shows the number of daily infections.

\begin{figure}[H]
    \centering
    \includegraphics[width=1.0\textwidth]{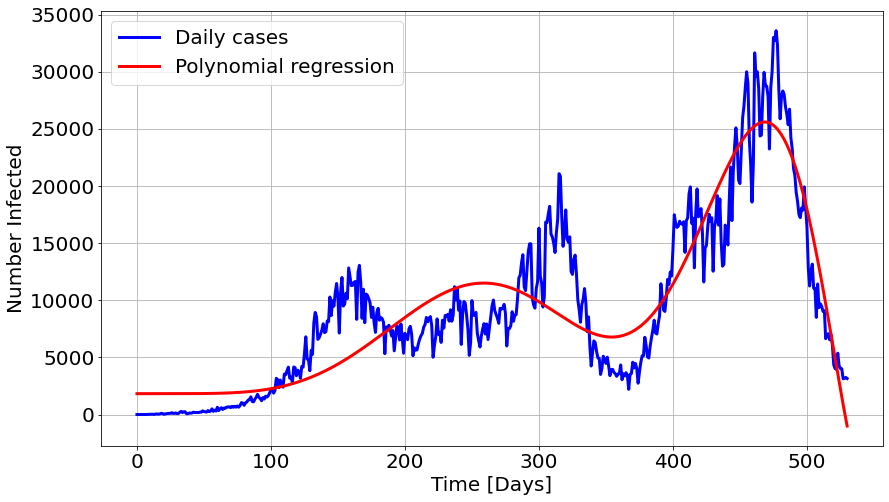}
    \caption{Graph of the fit made by the Machine Learning polynomial regression method for the daily infection data, with a polynomial of degree ten (10). The red curve indicates the adjustment made and the blue curve is the real data of daily infected by the COVID-19 disease.}
    \label{fig_6}
\end{figure}

\noindent The first approximation is the number of daily infections, which was to adjust to model this profile using the Machine Learning Polynomial Regression method. The result obtained in this adjustment can be seen in Figure \ref{fig_6}.

\begin{figure}[H]
    \centering
    \includegraphics[width=0.8\textwidth]{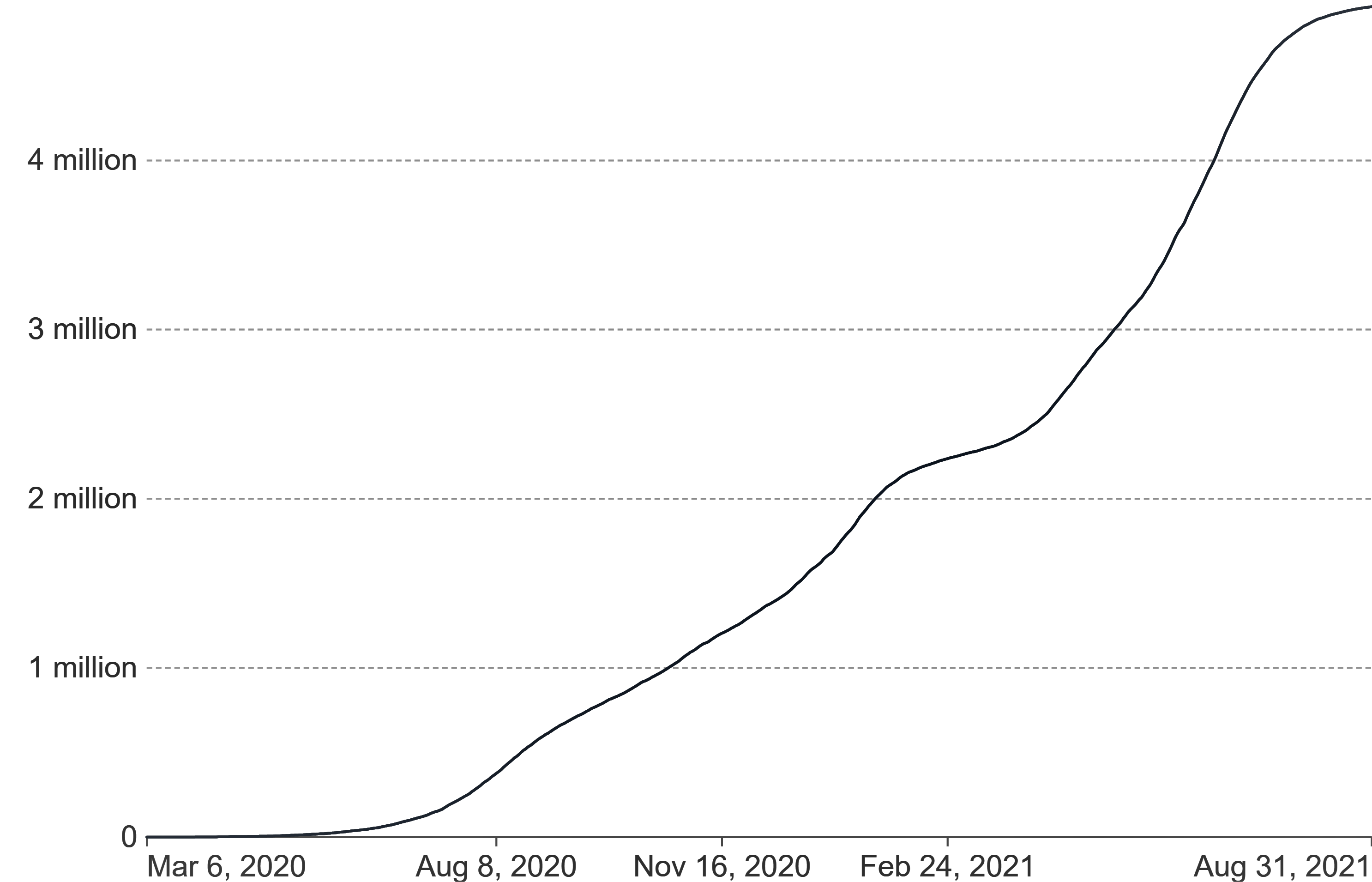}
    \caption{Graph of the accumulated number of confirmed daily cases for the infection of the COVID-19 disease in Colombia.}
    \label{fig_7}
\end{figure}

\noindent Due to the high dispersion of the data regarding the report of daily infections by COVID-19 in Colombia, applying any numerical adjustment method to this type of profile, such as the one shown in Figure \ref{fig_6}, is not recommended, and neither is it an optimal performance of the implemented mathematical method will be obtained. For this reason, analyzing the contagion data accumulated daily is more efficient, as shown in Figure \ref{fig_7}.

\begin{figure}[H]
    \centering
    \includegraphics[width=0.8\textwidth]{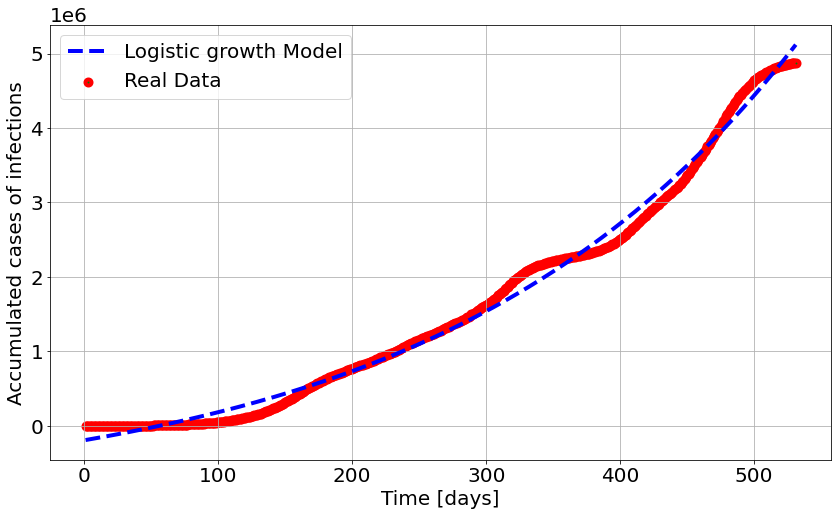}
    \caption{Graph of the number of infected cases as a function of time in days. The continuous curve in red shows the real data taken from the database of the Ministry of Health of Colombia, while the discontinuous curve in blue shows the adjustment made with a logistic growth function.}
    \label{fig_8}
\end{figure}

\noindent For Figure \ref{fig_7}, the implementation of the adjustment method with an increasing logistic function will be carried out through a logistic growth model to model the number of infections accumulated daily due to the COVID-19 disease in Colombia. In Figure \ref{fig_8}, the adjustment achieved is shown through the logistic growth method, with which it was possible to represent a function capable of modeling the number of accumulated cases infected by the COVID-19 disease.

\begin{figure}[H]
    \centering
    \includegraphics[width=0.8\textwidth]{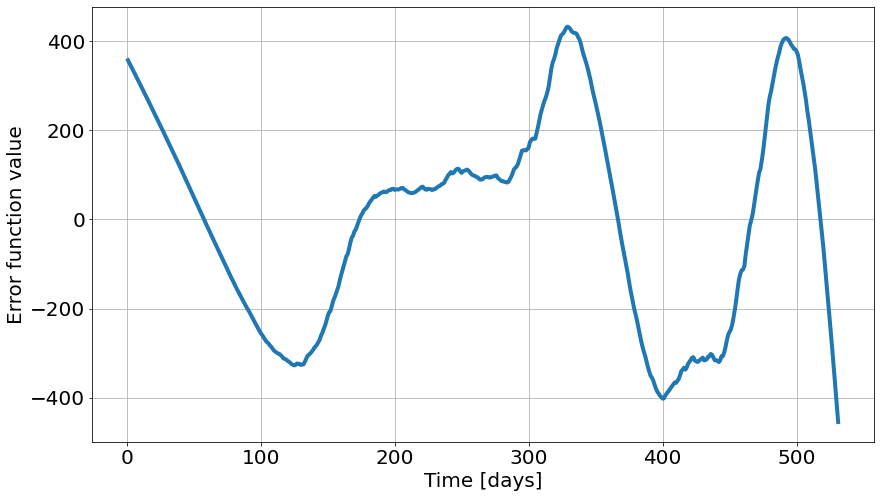}
    \caption{The error function shows the difference between the real data and the logistic growth model for the total number of accumulated daily cases of people infected by COVID-19 in Colombia.}
    \label{fig_9}
\end{figure}

\noindent On the other hand, in Figure \ref{fig_8}, an outstanding representation of the logistic growth model can be seen for the data obtained from the data of accumulated cases of daily infection in Colombia. In order to observe this difference in terms of this model obtained with the real data in Figure \ref{fig_8}, the difference between the values obtained between the points of the logistic growth model and the real data was calculated, which leads us to perform an error function, to be able to evaluate how efficient the model is for this set of real data taken in Colombia (See Figure \ref{fig_9}).

\begin{figure}[H]
    \centering
    \includegraphics[width=0.8\textwidth]{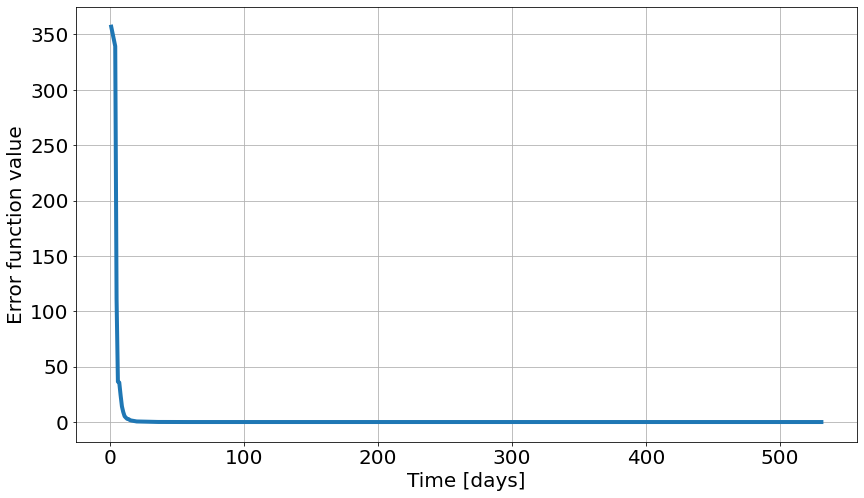}
    \caption{The second error function, the difference between the real data and the logistic growth model weighed with the real data by the total number of days for the total number of accumulated daily cases of people infected by COVID-19 in Colombia.}
    \label{fig_10}
\end{figure}

\noindent A second possible error function to present and show the efficiency of the function obtained from the logistic growth model can be seen in Figure \ref{fig_10}, where the difference between the model and the real data was made and said the difference was weighed. With the product of the real data by the total number of days in which this record of values was made (See Figure \ref{fig_10}).

\begin{figure}[H]
    \centering
    \includegraphics[width=0.8\textwidth]{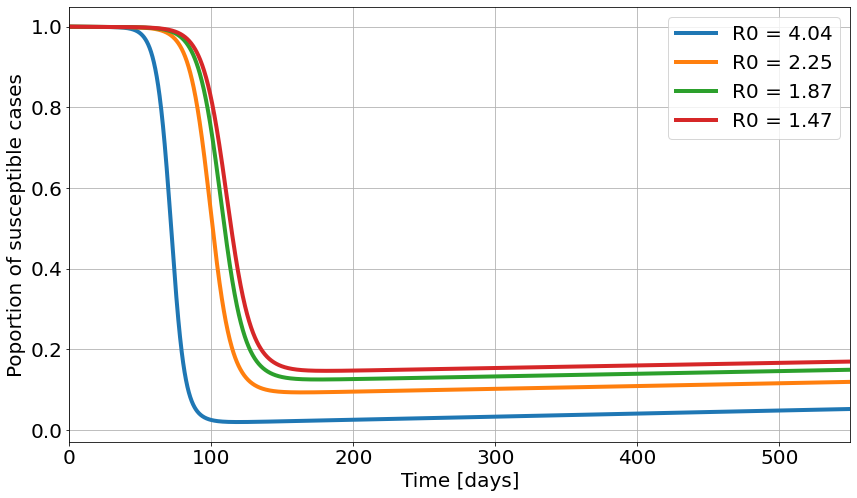}
    \caption{Proportion of the population susceptible to being infected with the SARS-CoV-2 virus, during the period studied in this research (550 days), for four (4) different prevention and protection measures against the virus.}
    \label{fig_11}
\end{figure}

\noindent Continuing with the analysis of the results carried out in this research project, we will now continue to review the modified SEIR model proposed in the theoretical framework of this article, where we will begin by analyzing the curves obtained, for the number of people Susceptible, to being infected with the SARS-CoV-2 virus. In Figure \ref{fig_11}, the different profiles of the curves for the cases susceptible to becoming infected with the COVID-19 disease were shown, taking into account four different scenarios.

\begin{figure}[H]
    \centering
    \includegraphics[width=0.8\textwidth]{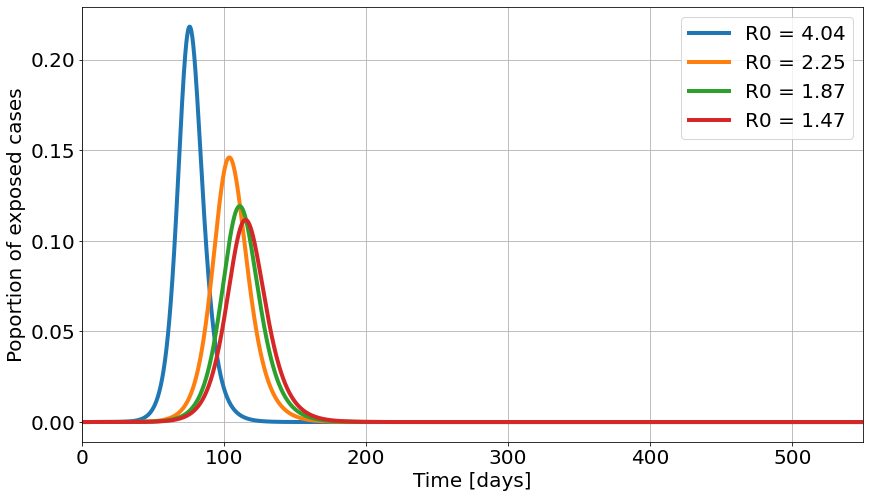}
    \caption{The proportion of people exposed to the virus, during the period studied in this research (550 days), for four (4) different prevention and protection measures against the virus.}
    \label{fig_12}
\end{figure}

\noindent Another group or category of people in the middle of a pandemic studied in the SEIR model; is the proportion of people exposed to being infected by the virus; Figure \ref{fig_12} shows the behavior of this group of people for four (4) different scenarios, in terms of control and applied care measures shown in the value of $R0$.

\begin{figure}[H]
    \centering
    \includegraphics[width=0.8\textwidth]{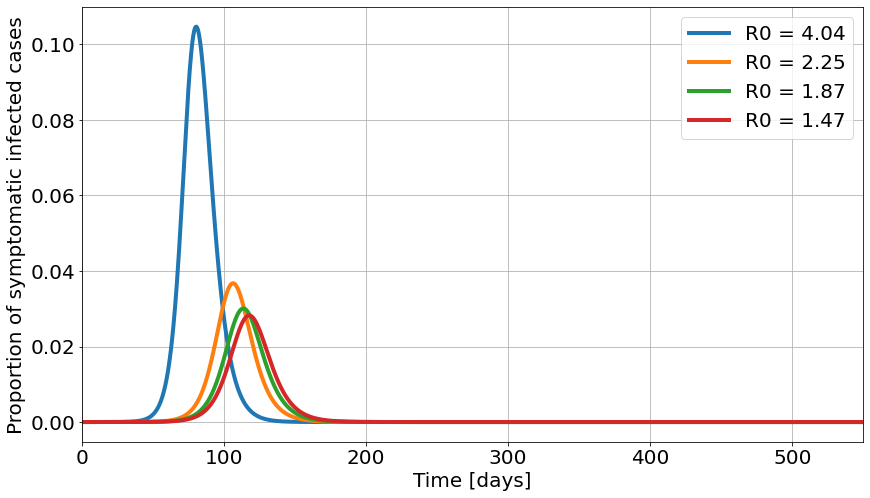}
    \caption{The proportion of symptomatic people infected with the virus, during the studied period (550 days), for four (4) different prevention and protection measures against the virus.}
    \label{fig_13}
\end{figure}

\noindent Continuing with the analysis of the proportion of cases of people who contracted the disease and who present symptoms (symptomatic infected), it can be seen in Figure \ref{fig_13}, a decrease by half compared to the peak of sensitive cases (See Figure \ref{fig_12}), for the value of $ R0 = 4.04 $.

\noindent On the other hand, it can be seen that the width of peaks of the Gaussian bells, shown in figures \ref{fig_12} and \ref{fig_13}, remain constant and symmetrical for the four scenarios. On the other hand, it can be seen how in these two figures, it is expected that in an approximate time of 170 days, for a value of $R0 = 1.47$, from the beginning of the study of the pandemic, the symptomatic infected and susceptible cases will be almost null (no cases are recorded).

\begin{figure}[H]
    \centering
    \includegraphics[width=0.8\textwidth]{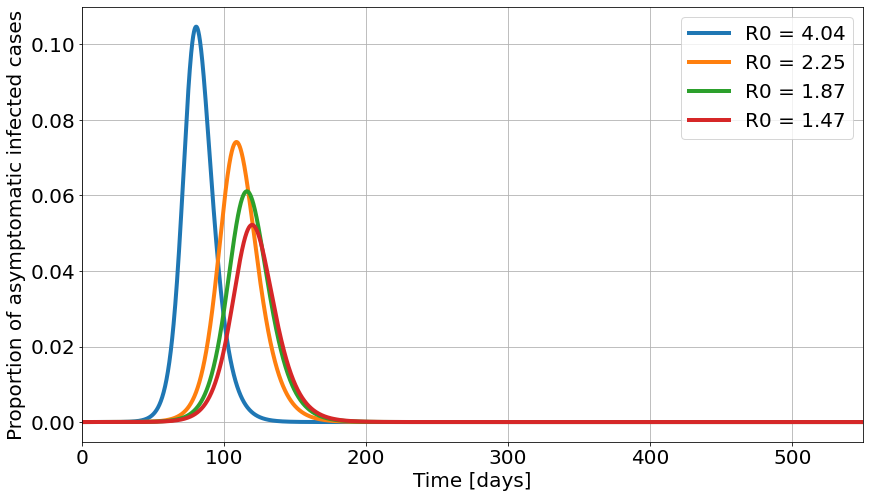}
    \caption{The proportion of people asymptomatic infected with the virus, during the studied period (550 days), for four different prevention and protection measures against the virus.}
    \label{fig_14}
\end{figure}

\noindent In Figure \ref{fig_14}, the proportion of asymptomatic infected people is shown; it can also be seen how the infection peaks in this figure are higher than in the case of the symptomatic infected shown in Figure \ref{fig_13}, which shows lower peaks of the proportion of infected people. It is for this reason that care and prevention measures against the spread of the COVID-19 disease must always be maintained, regardless of the people around us or in what space we are, since it can be evidenced that a more significant number of infected people are asymptomatic; compared to the number of people who contract the virus and are symptomatic (See Figures \ref{fig_13} and \ref{fig_14}).

\begin{figure}[H]
    \centering
    \includegraphics[width=0.8\textwidth]{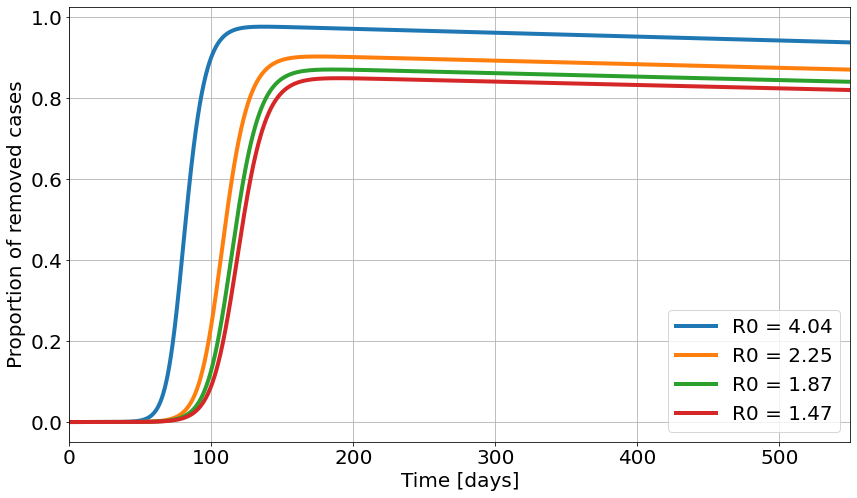}
    \caption{The proportion of people removed, during the studied period (550 days), for four different prevention and protection measures against the virus.}
    \label{fig_15}
\end{figure}

\noindent The proportion of cases removed (See Figure \ref{fig_15}), in those cases of people who recovered from the disease and who unfortunately died after being infected by the virus SARS-CoV-2.

\noindent In Figure \ref{fig_15}, there is an increasing behavior in the proportion of cases removed between the first 50 and 150 days where the pandemic was studied.

\noindent 

\section{Conclusions}
\label{sec_4}

In this work, the simulation of four scenarios was carried out, taking into account different $R0$ parameters, which can represent and reproduce different prevention scenarios, plans, and strategies against the spread and infection of the SARS-CoV-2 virus.

\noindent Further in this article, the results obtained by different models, reproduced by the methods implemented, were shown, within which an analysis was made of the number of new infections by COVID-19 disease, for which was used and implemented the Polynomial Regression Machine Learning method.

\noindent The logistic growth method was also used, with which it was possible to obtain an adjustment model for the number of new accumulated infections of the COVID-19 disease, during the time studied throughout the pandemic. For the application of this method, the error functions obtained were presented from the real contagion data in Colombia.

\noindent The results obtained from the simulation carried out to model four different scenarios are presented, which describe the development and behavior of the pandemic shown from those four different possible situations in a defined period.

\section*{Acknowledgements}

The author: Alex Estupiñán, would like to express their thanks, especially to the Universidad de Investigación y Desarrollo UDI, for all the human, material and financial support to carry out this research work. 



\end{document}